\documentclass[conference]{IEEEtran}
\IEEEoverridecommandlockouts
\usepackage{cite}
\usepackage{amsmath,amssymb,amsfonts}
\usepackage{algorithmic}
\usepackage{graphicx}
\usepackage{textcomp}
\usepackage{subfigure}
\usepackage{xcolor}
\usepackage{soul}
\usepackage{diagbox} 
\usepackage{multirow}
\usepackage[ruled]{algorithm2e}
\usepackage{color}
 \usepackage[normalem]{ulem}
\usepackage{marvosym}
 \useunder{\uline}{\ul}{}
\newcommand{\nop}[1]{}

\usepackage{colortbl}  

\def\BibTeX{{\rm B\kern-.05em{\sc i\kern-.025em b}\kern-.08em
    T\kern-.1667em\lower.7ex\hbox{E}\kern-.125emX}}
\begin{document}

\title{Group Buying Recommendation Model\\ Based on Multi-task Learning\thanks{Corresponding author is Deqing Yang. This work was supported by Shanghai Science and
Technology Innovation Action Plan No.21511100401.}}
\author{\IEEEauthorblockN{Shuoyao Zhai$^1$$^{\S}$,
Baichuan Liu$^1$$^{\dag}$,
Deqing Yang$^1$$^3$$^{\dag}$\textsuperscript{\Letter},
Yanghua Xiao$^2$$^3$$^{\dag}$}
\IEEEauthorblockA{$^1$School of Data Science, Fudan University, Shanghai 200433, China}
\IEEEauthorblockA{$^2$School of Computer Science, Fudan University, Shanghai 200433, China}
\IEEEauthorblockA{$^3$Shanghai Key Laboratory of Data Science, Shanghai 200433, China}
\IEEEauthorblockA{$^{\S}$sydi21@m.fudan.edu.cn, $^{\dag}$\{bcliu20, yangdeqing, shawyh\}@fudan.edu.cn}
}

\maketitle

\begin{abstract}
In recent years, group buying has become one popular kind of online shopping activities, thanks to its larger sales and lower unit price. Unfortunately, seldom research focuses on the recommendations specifically for group buying by now. Although some recommendation models have been proposed for group recommendation, they can not be directly used to achieve the real-world group buying recommendation, due to the essential difference between group recommendation and group buying recommendation. In this paper, we first formalize the task of group buying recommendation into two sub-tasks. Then, based on our insights into the correlations and interactions between the two sub-tasks, we propose a novel recommendation model for group buying, namely MGBR, which is built mainly with a multi-task learning module. To improve recommendation performance further, we devise some collaborative expert networks and adjusted gates in the multi-task learning module, to promote the information interaction between the two sub-tasks. Furthermore, we propose two auxiliary losses corresponding to the two sub-tasks, to refine the representation learning in our model. Our extensive experiments not only demonstrate that the augmented representations in our model result in better performance than previous recommendation models, but also justify the impacts of the specially designed components in our model. To reproduce our model's recommendation results conveniently, we have provided our model's source code and dataset on \emph{https://github.com/DeqingYang/MGBR}.
\end{abstract}

\section{Introduction}\label{sec:intro}
In the last decade, online shopping has become one kind of frequent user activities on Web, and produced great value for many enterprises and society. More recently, 
more and more users would like to participate in a \emph{group buying} rather than commit an individual purchase for their favorite products, since one product's deal price in group buying is generally lower than that in individual purchase. 

\begin{figure}[t]
\hspace{0.2cm}
\includegraphics[width=0.9\linewidth]{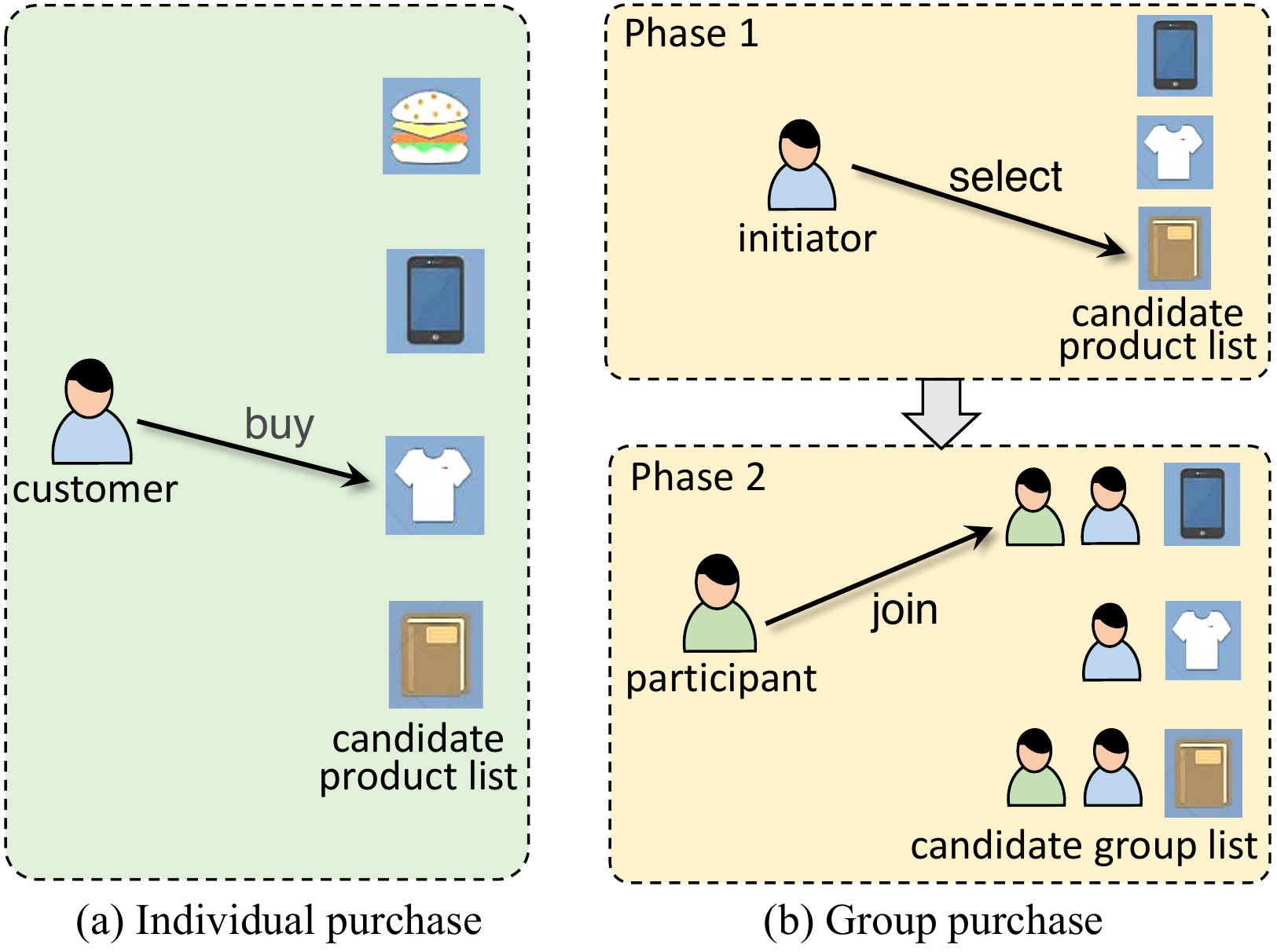}
\vspace{-0.2cm}
\caption{The comparison between individual purchase and group buying. In (a) individual purchase, a customer selects his/her favorite product from the candidate product list, to buy it without considering other users' purchases. In (b) group buying, an initiator selects one of his/her favorite products to launch a group buying at first, and waits for other users to participate in the group. Then, another user selects a group from the candidate group list, to join it as the role of participant.
}\label{fig:example}
\vspace{-0.3cm}
\end{figure}

Besides the deal price, a group buying is essentially different from an individual purchase. We illustrate it with the example in Fig. \ref{fig:example}, which comes from the real-world group buying process of online shopping. Fig. \ref{fig:example} (a) depicts the traditional individual purchase, where a user (customer) selects one favorite product from the candidate product list on the e-commerce platform, and directly buys it without considering other customers' purchases. By contrast, the group buying depicted by Fig. \ref{fig:example} (b) has two phases in fact. In the first phase, a user first selects one favorite product from the candidate list to launch a group for a more lower deal price. Such a user is identified as the \emph{initiator} of group buying. In the next phase, another user would select one group from the candidate group list, to join it as the role of \emph{participant}. Each candidate group consists of a product and the corresponding initiator, and may also include other participants who have joined the group before. Thus, there are three types of objects in one group, i.e., initiator, participant and item. A group buying is dealt only when the condition is satisfied, e.g., the number of participants satisfies the preset threshold.

Given the great value of recommender systems in promoting online shopping, the recommendation for group buying deserves being developed. Therefore, we dedicate to propose an effective recommendation model for group buying in this paper, which is seldom studied by previous research of recommendation. Similar to group recommendation \cite{Group1,Group2,Group}, there is also a user group consisting of one initiator and some participants in group buying, who interact with the same item. However, the existing models of group recommendation \cite{AttGroup,KgGroup,HierGroup,SocialGroup} can not be used directly to achieve group buying recommendation due to the following reasons.

 1. According to the two phases described in Fig. \ref{fig:example}, the recommendation task of group buying addressed in this paper can be divided into two sub-tasks. The first sub-task is to recommend items for a given initiator that he/she would like to launch a group buying. The second sub-task is to recommend participants to join the group given an initiator and the selected item. By contrast, the only task of group recommendation is just to recommend items for a given user group that the users in the group would like to buy. Accordingly, most of the existing group recommendation models mainly focus on how to learn group (preference) representations through aggregating the preferences of group members \cite{baltrunas2010group,AttGroup}.

 2. In addition, unlike group buying, the users in group recommendation do not play different roles of initiator and participant. The previous group recommendation models generally leverage the interactions of group-item-user to learn group representations \cite{SocialGroup,AttGroup}, and some models also leverage the relationships between different groups \cite{HierGroup}. Comparatively, the correlations and interactions between initiators, participants and items in group buying are more complicated. It is crucial for group buying modeling to leverage these complicated correlations and interactions sufficiently and correctly.


Therefore, it is challenging to design a recommendation model specifically for group buying. To achieve effective recommendation towards above distinct properties of group buying, we propose a novel recommendation model to accomplish the two sub-tasks of group buying recommendation simultaneously. We denote our model as \textbf{MGBR} (\textbf{M}uti-task learning based \textbf{G}roup \textbf{B}uying \textbf{R}ecommendation), in which the major multi-task learning module is harnessed to better learn the embeddings of initiators, items and participants. Specifically, based on our insights that the two sub-tasks of group buying recommendation are correlated to each other, we devise some expert networks along with adjusted gates in the multi-task learning module, to promote the information interaction between the two sub-tasks. As a result, the significant information is encoded into the learned embeddings to enhance our model's recommendation performance. In addition, in order to refine the representation learning in our model, two auxiliary losses corresponding to the two sub-tasks are further proposed to optimize our model training, resulting in more performance gains.

In summary, our contributions in this paper include:

 1. We formalize the task of group buying recommendation into two correlated sub-tasks, which indeed conforms to the real-world settings of online shopping. To the best of our knowledge, this is the first to propose a solution of accomplishing these two sub-tasks simultaneously. 

 2. We propose a novel recommendation model MGBR to achieve the objectives of group buying recommendation, in which we build a multi-task learning module and further devise adjusted gate units, to promote effective information interaction between the two sub-tasks.

 3. We further propose two auxiliary learning objectives based on our deep insights into the two sub-tasks of group buying, to refine the representation learning in our model, resulting in further enhanced performance.

 4. Our extensive experiments upon a real-world group buying dataset not only demonstrate our model's advantage over the compared recommendation models, but also justify the impacts of the significant components in our model.

The rest of this paper is organized as follows. In Section \ref{sec:method}, we first formalize the task of group buying recommendation and then describe our model in detail. Next, we display the experiment results to verify our model's performance on group buying recommendation in Section \ref{sec:exp}. We introduce the representative works related to our research in Section \ref{sec:related} and conclude our work in Section \ref{sec:conclude}.

\section{Methodology}\label{sec:method}
In this section, we first formalize the recommendation task of group buying addressed in this paper, and then detail our proposed recommendation model for group buying.

\begin{figure*}[t]
  \centering
  \includegraphics[width=1\linewidth]{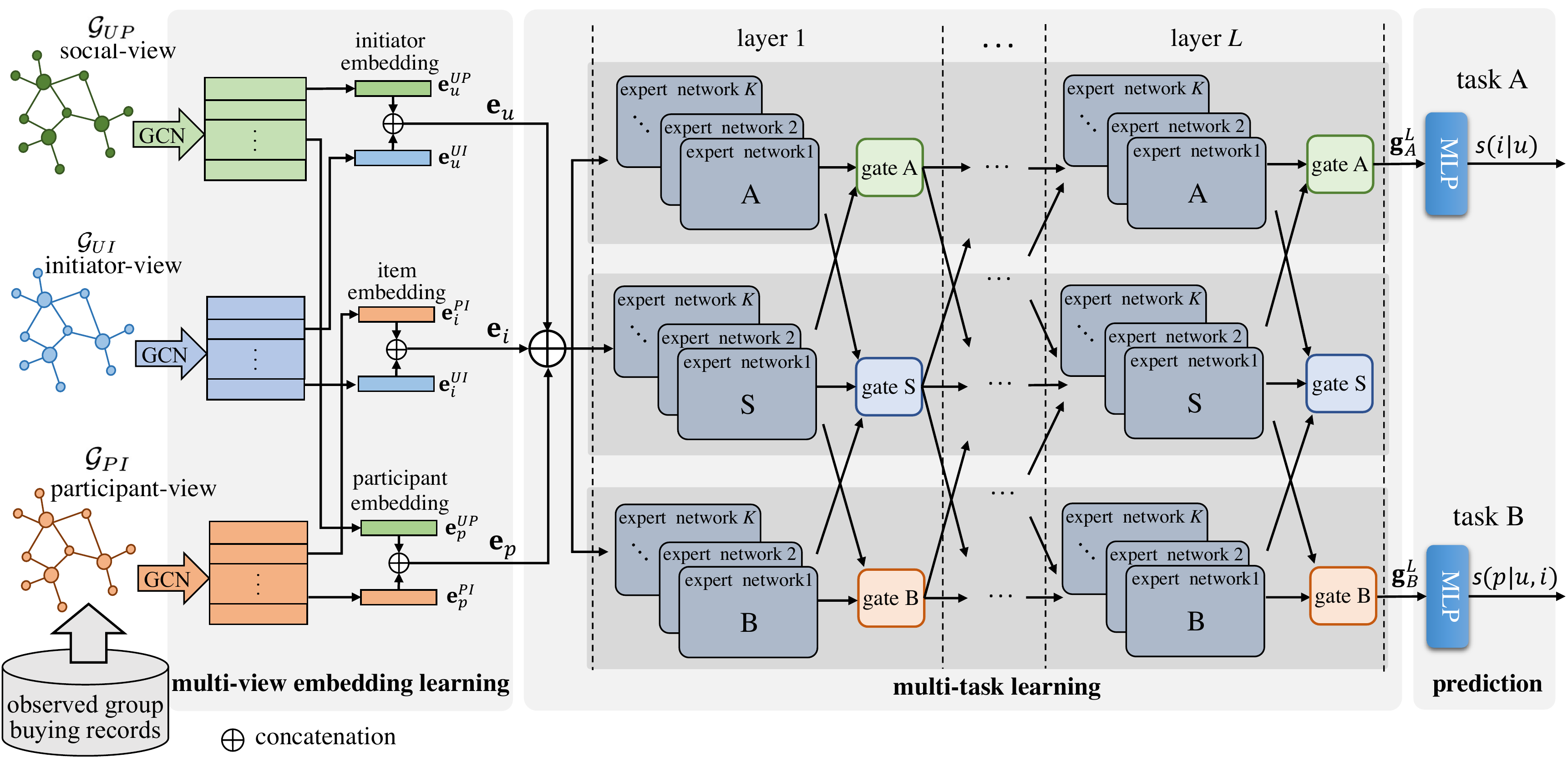}
  \caption{The overview of our proposed MGBR model for group buying recommendation. It consists of three major modules: multi-view embedding learning module, multi-task learning module and prediction module. In the first module, the embeddings of initiators, items and participants are learned through the GCNs on three views (graphs). The second module is built with some expert networks and gates, to further learn the object embeddings based on the information interaction between the two sub-tasks. In the prediction module, the prediction scores for candidate items and participants are respectively computed by two MLPs fed with the output embeddings in the previous module. }\label{fig:framework}
\end{figure*}

\subsection{Task Formalization}\label{sec:task}


At first, we use $u$, $i$, $p$ to represent an initiator, item and participant, respectively. According to our description about the group buying process, the two sub-tasks of group buying recommendation can be formalized as follows.

\textbf{Task A}: Recommending an $i$ for a given $u$ to launch a group buying through computing a scoring function $s(i|u)$.

\textbf{Task B}: Recommending a $p$ for a given pair $(u,i)$ through computing a scoring function $s(p|u,i)$.

As other recommendation models, the recommended objects ($i$ and $p$) are determined based on the computed $s(i|u)$ and $s(p|u,i)$ for all items ($i$) and participants ($p$) in the candidate lists, respectively.

Then, we explain the rationality of computing $s(i|u)$ and $s(p|u,i)$ to achieve the two sub-tasks of group buying recommendation. Formally, given an observed deal group $<u, i, G>$ where $G=\{p_1, p_2, ..., p_{|G|}\}$ is the participant group, a group buying recommendation model should maximize the probability of observing this deal group, that is $P(u, i, p_1,p_2, ..., p_{|G|})$. In most of group buying's real scenarios, the participants in one group are not familiar with each other, especially for the deal groups needs hundreds of participants. It indicates that each participant is independent to each other. Thus, we have
$$
\begin{aligned}
P(u, i, p_1, p_2, ..., p_{|G|}) = P(p_1, p_2, ..., p_{|G|}|u, i)P(i|u)P(u) \\
P(u, i, p_1, p_2, ..., p_{|G|}) \propto P(p_1, p_2, ..., p_{|G|}|u, i) P(i|u)\\
P(p_1, p_2, ..., p_{|G|}|u, i) = P(p_1|u, i)P(p_2|u, i)...P(p_{|G|}|u, i).\\
\end{aligned}
$$
Accordingly, the probability $P(i|u)$ and $P(p|u, i)$ should be maximized by a good recommendation model, which just correspond to (are proportional to) $s(i|u)$ and $s(p|u,i)$, respectively. 

The training set of our model learning is collected from the observed deal groups. Specifically, for an observed group $<u, i, G>$, $(u, i)$ is a positive sample of task A. And $(u, i, p_1), (u, i, p_2), ..., (u, i, p_{|G|})$ are the positive samples of task B. For task A and task B, their negative samples are generated through negative sampling, of which the details will be introduced in the following experiment section.

In addition, we denote all user set and item set as $\mathcal{U}$ and $\mathcal{I}$, respectively. Obviously, $u, p\in\mathcal{U}$ and $i\in\mathcal{I}$. We further use $\mathcal{N}^+$ and $\mathcal{N}^-$ to denote the positive samples and negative samples of model training, respectively.



\subsection{Model Overview}
We first introduce the pipeline (framework) of our model briefly, as depicted in Fig. \ref{fig:framework}, which consists of three major steps (modules), i.e., multi-view embedding learning, multi-task learning, and model prediction. At first, in the multi-view embedding learning module, the embeddings of all user (initiators and participants) and items are learned through the graph convolutional networks (GCNs) on three graphs (views) that include different types of connections between initiators, items and participants. The second module is a multi-task learning framework, which is built with some expert networks and gates to exploit the information interaction between the two sub-tasks, since the two sub-tasks are mutually correlated according to the primary principle of group buying. In model prediction module, a multi-layer perceptron (MLP) is built for each sub-task, which is fed with the embeddings output by the gates in the previous module, to computes the score for each candidate item or participant. Furthermore, in order to refine the representation learning of initiators, items and participants in our model, two auxiliary losses are proposed for the two sub-tasks to optimize our model training, resulting in more performance gains.

\subsection{Multi-view Embedding Learning with GCNs}
\subsubsection{Insights into Object Interactions among Groups}
As introduced in Section \ref{sec:task}, our model aims to compute $s(i|u)$ and $s(p|u,i)$, which are computed based on the embeddings of the three types of objects, i.e., $u$, $i$ and $p$. Thus, learning object embeddings is the preliminary step of our model. According to the positive samples from the observed deal groups, the massive connections (interactions) between the three types of objects are witnessed, based on which the embedding features can be extracted. For example, the user-item interaction graph in many collaborative-filtering (CF) based recommendation models \cite{lightGCN,NGCF} is often leveraged to encode the collaborative signals into the embeddings of users and items, which are significant for the model to generate accurate results. In fact, three undirected graphs can be constructed in our scenario of group buying recommendation, each of which includes two types of objects, i.e., $(u, i)$, $(p, i)$ and $(u, p)$. We denote these graphs as $\mathcal{G}_{UI}$, $\mathcal{G}_{PI}$ and $\mathcal{G}_{UP}$, and name them as \emph{initiator-view}, \emph{participant-view} and \emph{social-view}, respectively. All of the structural features in $\mathcal{G}_{UI}$, $\mathcal{G}_{PI}$ and $\mathcal{G}_{UP}$ are significant to learn the embeddings of $u$, $i$ and $p$. Given that GCN \cite{GCN}'s capability of modeling the connections between nodes in a graph, we also apply GCNs on the three graphs to learn their node embeddings. Then, the embeddings of $u$, $i$ and $p$ are obtained, and therefore we call this step as multi-view embedding learning. 

\subsubsection{Detailed Operations}
Specifically, $\mathcal{G}_{UI}$ only includes initiator nodes and item nodes. The edge between $u$ and $i$ is established in $\mathcal{G}_{UI}$ if $u$ selects $i$ to launch a group buying. In fact, $\mathcal{G}_{UI}$ is just the user-item interaction graph in task A. Similarly, $\mathcal{G}_{PI}$ only includes participant nodes and item nodes, and the edge between $p$ and $i$ is established if $p$ has joined the group buying involving $i$. For $\mathcal{G}_{UP}$, it only has initiator nodes and participant nodes, i.e., only has user nodes. The edge between $u$ and $p$ is established if $p$ has joined the group launched by $u$. Please note that, to save computation costs and reduce connection noises, the edge between any two participants of a group is omitted in $\mathcal{G}_{UP}$\footnote{We have verified that the variant of incorporating the edges between participants even has slightly poor performance.}. It is because we believe that in one group the initiator's preference is more important than other participants' preferences on learning a participant's preference, since a participant is often less aware who has also participated in the group. Thus, we learn a participant's preference from $\mathcal{G}_{UP}$ without $p$-$p$ edges.

In brief, $\mathcal{G}_{UI}$, $\mathcal{G}_{PI}$ and $\mathcal{G}_{UP}$ represent distinct connections between $u, i, p$. Particularly, user preferences on items can be extracted from $\mathcal{G}_{UI}$ and $\mathcal{G}_{PI}$, but they focus on different user roles, i.e., initiator and participant. The preference similarities between users are extracted from $\mathcal{G}_{UP}$, which help the model better recommend participants for a given group.


Formally, suppose $\mathbf{X}^l_{UI}\in\mathbb{R}^{(|\mathcal{U}|+|\mathcal{I}|)\times d}$, $\mathbf{X}^l_{PI}\in\mathbb{R}^{(|\mathcal{U}|+|\mathcal{I}|)\times d}$ and  $\mathbf{X}^l_{UP}\in\mathbb{R}^{|\mathcal{U}|\times d}$ are the node embedding matrices learned at the $l$-th $(1\leq l\leq H)$ layer of the GCN on $\mathcal{G}_{UI}$, $\mathcal{G}_{PI}$ and $\mathcal{G}_{UP}$, respectively, and $d$ is embedding dimension. Then, the computations on the $l$-th layer are as follows,
\begin{equation} \mathbf{X}^{l}_{UI} = \sigma(\hat{A}_{UI}\mathbf{X}^{l-1}_{UI}\mathbf{W}^{l-1}_{UI}), \end{equation} 
\begin{equation} \mathbf{X}^{l}_{PI} = \sigma(\hat{A}_{PI}\mathbf{X}^{l-1}_{PI}\mathbf{W}^{l-1}_{PI}), \end{equation}
\begin{equation} \mathbf{X}^{l}_{UP} = \sigma(\hat{A}_{UP}\mathbf{X}^{l-1}_{UP}\mathbf{W}^{l-1}_{UP}), \end{equation} 
where $\hat{A}_{UI}\in\mathbb{R}^{(|\mathcal{U}|+|\mathcal{I}|)\times(|\mathcal{U}|+|\mathcal{I}|)}$, $\hat{A}_{PI}\in\mathbb{R}^{(|\mathcal{U}|+|\mathcal{I}|)\times(|\mathcal{U}|+|\mathcal{I}|)}$ and $\hat{A}_{UP}\in\mathbb{R}^{|\mathcal{U}|\times|\mathcal{U}|}$ are the normalized adjacency matrices with self-loops of $\mathcal{G}_{UI}$, $\mathcal{G}_{PI}$ and $\mathcal{G}_{UP}$, respectively. And $\mathbf{W}^{l-1}_{UI}, \mathbf{W}^{l-1}_{PI}, \mathbf{W}^{l-1}_{UP}\in\mathbb{R}^{d\times d}$ are trainable parameter matrices, $\sigma$ is Sigmoid function. Particularly, $\mathbf{X}^0_{UI},\mathbf{X}^0_{PI},\mathbf{X}^0_{UP}$ are randomly initialized by a Gaussian distribution (0, 1).

As shown in Fig. \ref{fig:framework}, each of $u, i, p$ has two embeddings of different views through the GCNs on the three views, since each of them is involved in two views.
As a result, we directly obtain the multi-view embeddings of $u, i, p$ as follows,
\begin{equation} \mathbf{e}_u = \mathbf{e}^{UI}_u || \mathbf{e}^{UP}_u, \end{equation} 
\begin{equation} \mathbf{e}_i = \mathbf{e}^{UI}_i || \mathbf{e}^{PI}_i, \end{equation} 
\begin{equation} \mathbf{e}_p = \mathbf{e}^{PI}_p || \mathbf{e}^{UP}_p, \end{equation} 
where $||$ is concatenation operator, and $\mathbf{e}^{UI}_u, \mathbf{e}^{UP}_u\in\mathbb{R}^d$ are $u$'s embeddings output by the GCN's $H$-th layer on $\mathcal{G}_{UI}$ and $\mathcal{G}_{UP}$, perspectively. The similar notations are used for $i$ and $p$. The $\mathbf{e}_u, \mathbf{e}_i, \mathbf{e}_p\in\mathbb{R}^{2d}$ learned in this module will be used as the input of the next multi-task learning module.

\begin{figure*}[t]
  \centering
  \includegraphics[width=0.81\textwidth]{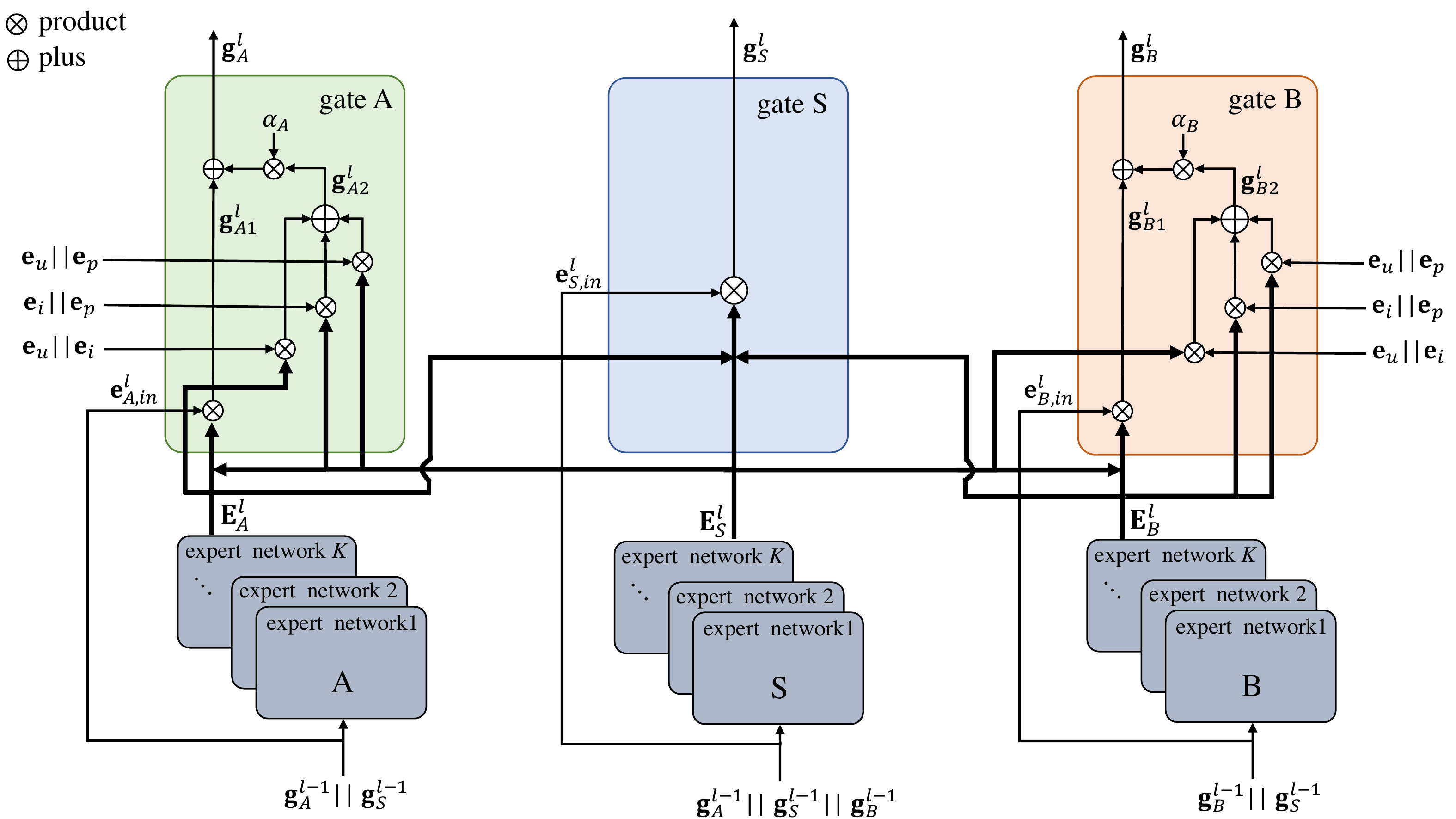}
  \caption{The architecture of the $l$-th layer in the multi-task learning module of our model. It consists of $K$ expert networks and one gate corresponding to task A, task B and shared information S, respectively.}\label{fig:gate}
\end{figure*}

\subsection{Multi-task Learning with Expert Networks and Gated Units}
\subsubsection{Insights into Learning Two Sub-tasks}\label{sec:intui}
It is reasonable to leverage multi-task learning framework in our model, since the objective of group buying recommendation are divide into task A and task B, as we mentioned before. We should notice that these two sub-tasks are mutual correlated. In task A, user $u$ selects item $i$ from the candidate item list to launch a group buying, not only based on his/her preference on $i$ but also the consideration whether $i$ is more favorable among the latent participants. It is up to the specific setting in real-wold group buying. In general, the more latent participants would like to buy $i$, the more likely the group buying of $i$ is to be achieved. Recall the example in Fig. \ref{fig:example} (b), the initiator in Phase 1 may like all of the recommended cellphone, T-shirt and book. He would prefer to select the cellphone instead of the book to launch a group buying, if he knows in advance that two participants will follow him to join the group of buying the cellphone in Phase 2, who are more than the followers in other two groups.

Accordingly, our model's learning of task A requires the intervention of participant information (embeddings) in task B, which encode participants' preference features. In addition, the learning of task B also needs the information of initiator and item in task A, since the objective of task B is to compute $s(p|u, i)$. Notably, each pair in the triple $(u,i,p)$ indicates distinct information. Particularly, $(u,i)$ indicates information of $u$'s preferences on $i$, which is more focused on by task A. $(u,p)$ indicates the preference similarity between $u$ and $p$, and $(i,p)$ indicates the information of $p$'s preference on $i$, both of which are more focused on by task B. 


\subsubsection{Detailed Operations}
Inspired by above insights into learning two sub-tasks in group buying recommendation, some information encoded into the object embeddings should be shared sufficiently by task A and task B, to enhance the learning effects of two sub-tasks. Given that the expert network is an effective solution for multi-task learning \cite{ma2018modeling}, we also build some expert networks along with gates in this module to achieve learning two sub-tasks, of which the architecture is depicted in Fig. \ref{fig:gate}. In brief, there are three sub-modules in our multi-task learning module, which correspond to the learning of task A, task B and shared information (denoted as S), respectively. For each sub-module, $K$ expert networks collaborate with one gate to learn object embeddings.


For clear statement, we use $A, B, S$ to denote the three sub-modules, and distinguish their corresponding notations. Formally, suppose $\mathbf{e}_{A_{i}}^{l}\in\mathbb{R}^{d}$ is the output embedding of the $i$-th $(1\leq i\leq K)$ expert network A at layer $l$ $(1\leq l\leq L)$, and computed as
\begin{equation}\label{eq:Ea}
\mathbf{e}_{A_{i}}^{l}= (\mathbf{g}_{A}^{l-1} || \mathbf{g}_{S}^{l-1})\mathbf{W}_{A_{i}}^{l},
\end{equation}
where $\mathbf{g}_{A}^{l-1}, \mathbf{g}_{S}^{l-1}\in\mathbb{R}^{d}$ are the output embeddings of gate A and gate S at layer $l-1$, respectively, and $\mathbf{W}_{A_{i}}^{l-1}\in\mathbb{R}^{2d\times d}$ is a trainable weight matrix. 

Similarly, we also compute the output embedding of the $i$-th expert network B at layer $l$ as
\begin{equation}\label{eq:Eb}
\mathbf{e}_{B_{i}}^{l}= (\mathbf{g}_{B}^{l-1} || \mathbf{g}_{S}^{l-1} )\mathbf{W}_{B_{i}}^{l},
\end{equation}
where $\mathbf{g}_{B}^{l-1}\in\mathbb{R}^{d}$ is the output embedding of gate B at layer $l-1$, and $\mathbf{W}_{B_{i}}^{l-1}\in\mathbb{R}^{2d\times d}$ is a trainable weight matrix.

For the output embedding of the $i$-th expert network S at layer $l$, which should involve gate A and gate B besides gate S. Thus, we have
\begin{equation}\label{eq:Es}
\mathbf{e}_{S_{i}}^{l}= (\mathbf{g}_{A}^{l-1} || \mathbf{g}_{S}^{l-1}|| \mathbf{g}_{B}^{l-1} )\mathbf{W}_{S_{i}}^{l},
\end{equation}
where $\mathbf{W}_{S_{i}}^{l-1}\in\mathbb{R}^{3d\times d}$ is also a trainable weight matrix. 

Next, we introduce how to compute the output embeddings of all gates mentioned in Eq. \ref{eq:Ea} $\sim$ \ref{eq:Es} as follows. 

To fully reflect the insights introduced in Section \ref{sec:intui}, we append the adjusted gated unit for gate A, B and S,  besides the generic gated unit, which follows the principle of self-attention in \cite{tang2020progressive}. According to this principle, the embeddings learned by the gate are the attentive aggregation of the embeddings learned by expert network A, B and S. Specifically, the output embedding of each gate consists of two sections. 

For gate A, the first section of output embedding at layer $l$ is calculated through the following generic gated operations as
\begin{equation}\label{eq:gA1}
\begin{aligned}
&\mathbf{g}^l_{A1} = \mathbf{e}^l_{A,in}\mathbf{W}_A
\left[\begin{array}{c}
	\mathbf{E}^{l}_{A}\\
        \mathbf{E}^{l}_{S}
    \end{array}    
\right]\in\mathbb{R}^{d},\\
&\mathbf{E}^{l}_{A}=
\left[\begin{array}{c}
	\mathbf{e}^{l}_{A_1}\\
        \mathbf{e}^{l}_{A_2}\\
        \vdots\\
        \mathbf{e}^{l}_{A_K}\\       
    \end{array}    
\right],\quad
\mathbf{E}^{l}_{S}=
\left[\begin{array}{c}
	\mathbf{e}^{l}_{S_1}\\
        \mathbf{e}^{l}_{S_2}\\
        \vdots\\
        \mathbf{e}^{l}_{S_K}\\       
    \end{array}    
\right],\\
&\mathbf{e}^l_{A,in}=\mathbf{g}_{A}^{l-1} || \mathbf{g}_{S}^{l-1},
\end{aligned}
\end{equation}
where $\mathbf{e}^l_{A,in}\in\mathbb{R}^{2d}$ is the input embedding of expert network A at layer $l$, and $\mathbf{W}_A\in\mathbb{R}^{2d\times 2K}$ is a trainable weight matrix. Eq. \ref{eq:gA1} implies that the information of $\mathbf{g}^l_{A1}$ is extracted from the expert network at layer $l$ in terms of the information in the previous layer.

Furthermore, the second section of gate A's output embedding at layer $l$ is calculated as
\begin{equation}\label{eq:ga2}
\begin{split}
&\mathbf{g}^l_{A2}  = (\mathbf{e}_u||\mathbf{e}_i)\mathbf{W}_{A,ui}\mathbf{E}^{l}_{A} +(\mathbf{e}_i||\mathbf{e}_p)\mathbf{W}_{A,ip}\mathbf{E}^{l}_{S} \\
& \quad\quad\quad+ (\mathbf{e}_u||\mathbf{e}_p)\mathbf{W}_{A,up}\mathbf{E}^{l}_{S}, 
\end{split}
\end{equation}
where $\mathbf{W}_{A,ui}, \mathbf{W}_{A,ip},\mathbf{W}_{A,up}\in\mathbb{R}^{4d\times K}$ are both trainable weight matrices, and $\mathbf{g}^l_{A2}\in\mathbb{R}^d$. Eq. \ref{eq:ga2} implies that the information of $\mathbf{g}^l_{A2}$ is also extracted from the expert network at layer $l$ but in terms of the input $(u, i, p)$'s information. In addition, the output embeddings of expert network A interact with $\mathbf{e}_u||\mathbf{e}_i$ instead of $\mathbf{e}_i||\mathbf{e}_p$ and $\mathbf{e}_u||\mathbf{e}_p$, because task A focuses more on the information of $(u,i)$. Meanwhile, the information from $(i,p)$ and $(u,p)$ are transferred through expert network S.

Then, the overall output embedding of gate A at layer $l$ is
\begin{equation}\label{eq:gA}
\mathbf{g}^l_{A} = \mathbf{g}^l_{A1} + \alpha_A \mathbf{g}^l_{A2},
\end{equation}
where $\alpha_A\in(0,1)$ is a control coefficient. 

Similarly, the output embedding of gate B is calculated as
\begin{equation}\label{eq:gB}
\begin{aligned}
&\mathbf{g}^l_{B1} = \mathbf{e}^l_{B,in}\mathbf{W}_B 
\left[\begin{array}{c}
	\mathbf{E}^{l}_{B}\\
        \mathbf{E}^{l}_{S}
    \end{array}    
\right],\quad
\mathbf{E}^{l}_{B}=
\left[\begin{array}{c}
	\mathbf{e}^{l}_{B_1}\\
        \mathbf{e}^{l}_{B_2}\\
        \vdots\\
        \mathbf{e}^{l}_{B_K}\\       
    \end{array}    
\right],\\
&\mathbf{e}^l_{B,in}=\mathbf{g}_{B}^{l-1} || \mathbf{g}_{S}^{l-1},\\
&\mathbf{g}^l_{B2} 
 = (\mathbf{e}_u||\mathbf{e}_i)\mathbf{W}_{B,ui}\mathbf{E}^{l}_{S} + (\mathbf{e}_i||\mathbf{e}_p)\mathbf{W}_{B,ip}\mathbf{E}^{l}_{B} \\
& \quad\quad\quad+ (\mathbf{e}_u||\mathbf{e}_p)\mathbf{W}_{B,up}\mathbf{E}^{l}_{B},\\
&\mathbf{g}^l_{B} = \mathbf{g}^l_{B1} + \alpha_B \mathbf{g}^l_{B2}.
\end{aligned}
\end{equation}

Compared with $\mathbf{g}^l_{A}$ and $\mathbf{g}^l_{B}$, the computation of $\mathbf{g}^l_{S}$ is not so complicate, which is calculated as
\begin{equation}\label{eq:gs}
\begin{split}
\mathbf{g}^l_{S} = \mathbf{e}^l_{S,in}\mathbf{W}_S
 \left[\begin{array}{c}
	\mathbf{E}^{l}_{A}\\
        \mathbf{E}^{l}_{S}\\
        \mathbf{E}^{l}_{B}
    \end{array}    
\right], \quad\mathbf{e}^l_{S,in}=\mathbf{g}_{A}^{l-1} || \mathbf{g}_{S}^{l-1}|| \mathbf{g}_{B}^{l-1},
\end{split}
\end{equation}
where $\mathbf{W}_S\in\mathbb{R}^{3d\times 3K}$ is also a trainable weight matrix.

In addition, to compute $\mathbf{e}_{A_{i}}^{1}, \mathbf{e}_{B_{i}}^{1}, \mathbf{e}_{S_{i}}^{1}\in\mathbb{R}^d$ we set
\begin{equation}
\mathbf{g}^0_{A}= \mathbf{g}^0_{B}= \mathbf{g}^0_{S}=\mathbf{e}_u||\mathbf{e}_i||\mathbf{e}_p,
\end{equation}
which are just the inputs of the expert networks at the first layer, and also the inputs of the multi-task learning module. 
Accordingly, the weight matrices in the first layer of expert networks have different sizes to that in other layers, i.e., $\mathbf{W}^1_{A_i}\in\mathbb{R}^{6d \times d}, \mathbf{W}^1_{B_i}\in\mathbb{R}^{6d \times d}, \mathbf{W}^1_{S_i}\in\mathbb{R}^{9d \times d}$.

\subsection{Prediction Module}
In this module, our model predicts $s(i|u)$ and $s(p|u,i)$ respectively with the output embeddings of gate A and gate B at the final layer (layer $L$)  respectively. As we introduced before, the two scores are used to achieve the two sub-tasks in group buying recommendation, and computed as
\begin{equation}\label{eq:s1}
\begin{split}
&s(i|u)=\sigma(\operatorname{MLP_A}(\mathbf{g}_A^L))\\
&\quad\quad=\sigma\big(\operatorname{MLP_A}(\operatorname{MTL_A}(\mathbf{e}_u||\mathbf{e}_i||\mathbf{e}_p))\big),
\end{split}
\end{equation}
\begin{equation}\label{eq:s2}
\begin{split}
&s(p|u,i)=\sigma(\operatorname{MLP_B}(\mathbf{g}_B^L))\\
&\quad\quad=\sigma\big(\operatorname{MLP_B}(\operatorname{MTL_B}(\mathbf{e}_u||\mathbf{e}_i||\mathbf{e}_p))\big),
\end{split}
\end{equation}
where $\operatorname{MLP_{A/B}}()$ represents the operations of MLP fed with the output embedding of gate A/B at the last layer. $\operatorname{MTL_{A/B}}()$ represents the operations of our multi-task learning module. 
Then, the candidate item with the largest $s(i|u)$ is taken as the recommendation result of task A. Similarly, the candidate participant with the largest $s(p|u,i)$ is taken as the recommendation result of task B.

Please note that, the $\mathbf{e}_p$ in Eq. \ref{eq:s1} is obtained by a different way of that in Eq. \ref{eq:s2}. In task A, all users except for $u$ are the potential participants for the group buying of $i$. Thus, we average the embeddings of all users as the $\mathbf{e}_p$ in Eq. \ref{eq:s1}. By contrast, in task B, the $\mathbf{e}_p$ in Eq. \ref{eq:s2} is just the embedding of the candidate participant who needs being judged by the model.

\subsection{Model Optimization}
In fact, the recommendation task of our work is also the ranking-based recommendation of implicit feedback \cite{NCF,KTGAN,lightGCN,FMF,KAUR}. Thus, for both of the two sub-tasks in our model, we use the Bayesian ranking loss (BPR) \cite{BPR} as the objectives of our model optimization. Specifically, the overall loss of our model learning is as follows,
\begin{equation}\label{eq:L}
\mathcal{L}=\mathcal{L}_A +\beta\mathcal{L}_B,
\end{equation}
and 
\begin{equation}\label{eq:Lb}
\begin{split}
&\mathcal{L}_A=-\frac{1}{|\mathcal{N}^+_A\cup\mathcal{N}^-_A|}\sum\limits_{(u,i)\in\mathcal{N}^+_A}\sum\limits_{(u,i')\in\mathcal{N}^-_A}\log\sigma\big(s(i|u)-s(i'|u)\big),\\
&\mathcal{L}_B=-\frac{1}{|\mathcal{N}^+_B\cup\mathcal{N}^-_B|}\times\\
&\quad\quad\sum\limits_{(u,i,p)\in\mathcal{N}^+_B}\sum\limits_{(u,i,p')\in\mathcal{N}^-_B}\log\sigma\big(s(p|u,i)-s(p'|u,i)\big),
\end{split}
\end{equation}
where $\beta$ is the control coefficient (weight) of $\mathcal{L}_B$, $\mathcal{N}^+_{A/B}$ ($\mathcal{N}^-_{A/B}$) denotes the positive (negative) training samples of task A/B. We adopt negative sampling to collect the negative samples in our training set, which will be detailed in the subsequent experiment section.

In addition, we adopt Adam \cite{Adam} as the optimization method to train our model. 

\subsection{Refining Representation Learning with Auxiliary Losses}
Based on our insights into the two sub-tasks of group buying recommendation, we further propose two auxiliary losses for our model training to refine the representation (object embedding) learning in our model. These auxiliary losses' effects on enhancing recommendation performance are verified by our experiments.



\subsubsection{Task A's Auxiliary Loss}
Although the information of the participants in task B would influence task A, according to our aforementioned analysis, in task A the model should pay more attention to the match between $u$ and $i$, rather than the match between $(u,i)$ and $p$. It is because that the preference on $i$ is the primary reason for $u$ to launch a group buying with $i$. Another merit of such consideration is to avoid introducing more noise from the potential participants into task A learning. It implies that, for a positive (observed) triple $(u,i,p)$ from historical group buying records, replacing $i$ with other items degrades the model's prediction score more than replacing $p$ with other users. Accordingly, our model should try to ensure
\begin{equation}
\begin{aligned}
s(u,i,p) > s(u,i',p),\\
s(u,i,p') > s(u,i',p), 
\end{aligned}
\end{equation} 
where $s(u,i,p)$ is computed as $s(i|u)$ in Eq. \ref{eq:s1} except that $\mathbf{e}_p$ is just the embedding of the participant $p$. $i'$ ($p'$) is the corrupted item (participant) not existing in the true group buying records.

To propose the auxiliary loss for task A, for a given triple $t=(u,i,p)$ in the positive samples of task A's training set, we first define
$
\mathcal{T}_t^I=\{(u,i',p)|i'\in\mathcal{I}\setminus i\}
$ and
$
\mathcal{T}_t^P=\{(u,i,p')|p'\in\mathcal{U}\setminus G_{u,i}\}
$ where $G_{u,i}$ is the set of all participants who have joined the group including $u$ and $i$. In our experiments, we set $|\mathcal{T}_t^I|$ and $|\mathcal{T}_t^P|$ of any $t$ to the same fixed size, denoted as $|\mathcal{T}|$, through negative sampling. 

Besides BPR loss, the ListNet loss \cite{ListNet} is also a list-wise loss appropriate for ranking-based recommendation. Thus, based on above idea, we propose the following auxiliary loss of task A as
\begin{equation}\label{eq:La'}
\begin{aligned}
&\mathcal{L}'_A=-\frac{1}{|\mathcal{N}_A^+|\times 2|\mathcal{T}|}\times\\
&\quad\quad\quad\sum\limits_{t\in\mathcal{N}^+_A}\sum_{(u,i,p)\in \mathcal{T}_t^I\cup\mathcal{T}_t^P} y_{(u,i,p)}\log s(u,i,p),
\end{aligned}
\end{equation}
where $y_{(u,i,p)}=1$ if triple $(u,i,p)\in \mathcal{T}_p$, otherwise $y_{(u,i,p)}=0$. In fact, Eq. \ref{eq:La'} is proposed to fine-tune the embeddings of $u, i, p$, of which the idea is similar to contrastive learning \cite{CL}. 

\nop{
where $S()$ means the score obtained by our model, and $T_p(u,i,p)=(u,i,p'), p' \in P-P_{u,i}$ means replacing the $p$ in the original triple with the $p'$ obtained by negative sampling. Similarly, $T_{I}(u,i,p)=(u,i',p), i' \in I-\{i\}$ means replacing $i$ with the $i'$ obtained by negative sampling. When items change, users are likely to stop initiating group buyings, but when participant information changes, it does not affect the process of initiating group buyings. The above inequalities indicate that the changes of items should cause a greater negative impact, but the impact caused by participant should be minor.
}

\subsubsection{Task B's Auxiliary Loss}
The consideration of refining representation learning in task B is different to that in task A. On the one hand, replacing $i$ with other item $i'$ in a true triple $(u,i,p)$ should degrade the model's prediction score apparently, since $i'$ can not attract the participant to join the group due to the mismatch between $p$'s preference on $i'$. On the other hand, replacing $p$ with other user $p'$ in a true triple $(u,i,p)$ should also degrade the model's prediction score apparently, still due to the mismatch between $p'$'s preference on $i$. Accordingly, our model should try to ensure
\begin{equation}\label{eq:su}
s(p|u,i) > s(p|u,i'),
\end{equation}
and
\begin{equation}\label{eq:sp}
s(p|u,i) > s(p'|u,i).
\end{equation} 

In fact, encouraging Eq. \ref{eq:sp} is just the objective of task B, which has been reflected by the loss of Eq. \ref{eq:Lb}. Thus encouraging Eq. \ref{eq:su} can be regarded as the auxiliary objective of task B. Given that Eq. \ref{eq:su} is also a form of pair ranking, we implement it also by a BPR-based loss as
\begin{equation}
\begin{split}
&\mathcal{L}'_B=-\frac{1}{|\mathcal{N}_B^+|\times |\mathcal{T}|}\times\\
&\quad\quad\sum\limits_{t\in\mathcal{N}^+_B} \sum_{(u,i',p) \in \mathcal{T}_t^I} \log \sigma\big(s(p|u,i)-s(p|u,i')\big).
\end{split}
\end{equation}

Then, the overall loss of our model in Eq. \ref{eq:L} is modified as
\begin{equation}\label{eq:L1}
\mathcal{L}=\mathcal{L}_A +\beta\mathcal{L}_B+\beta_A\mathcal{L}'_A+\beta_B\mathcal{L}'_B,
\end{equation}
where $\beta_A, \beta_B$ are also the control coefficients of auxiliary losses.


\subsection{Time Complexity Analysis}
Given that we build multiple layers consisting of expert networks and gates in the multi-task learning module of MGBR, to achieve complicated information interaction between sub-module A, B and S, MGBR's time consumption is undoubtedly bigger than the recommendation models only with GCNs and the matrix factorization (MF) based models. We mainly analyze the time complexity of MGBR's multi-task learning module as follows, and display the empirical results of comparing the time consumption of MGBR and other models in the subsequent experiment section.

At first, the time complexity of multi-view embedding learning is $O\big(|\mathcal{U}|^2 d+ (|\mathcal{U}|+|\mathcal{I}|)^2 d+ (|\mathcal{U}|+|\mathcal{I}|)^2 d\big)=O(|\mathcal{U}|+|\mathcal{I}|)^2 d)$. Then, we analyze the time complexity of processing a sample $(u,i,p)$ in MGBR's multi-task learning module. For one layer consisting of $K$ expert networks and one gate, one expert consumes the time of $O(d^2)$, hence $3K$ experts consume $O(3Kd^2)=O(Kd^2)$. In the adjusted gate unit, the time complexity of computing $\mathbf{g}^l_{A1}$ in Eq. \ref{eq:gA1} is $O(2d\times 2K+d\times 2K)=O(Kd)$. And the time complexity of computing $\mathbf{g}^l_{A2}$ in Eq. \ref{eq:ga2} is $O\big(3(4d\times K+ d\times K)\big)=O(Kd)$. So it consumes $O(Kd)$ to obtain $\mathbf{g}^l_A$. Similarly, the calculation of $\mathbf{g}^l_B$ and $\mathbf{g}^l_S$ is also $O(Kd)$. Thus, the time complexity of $L$ layers in the multi-task learning module is $O\big(L(Kd^2+Kd+Kd)\big)=O(LKd^2)$. Given that $L$ and $K$ are generally small number, the time consumption of the multi-task learning module is approximate to $O(d^2)$, i.e., the square of embedding dimension, which is affordable.

\section{Experiments}\label{sec:exp}

In this section, we conduct extensive experiments on one real-world dataset of group buying to answer the following research questions:
\begin{itemize}
\item \textbf{RQ1}: Does our proposed recommendation model perform better than previous models on the recommendation of group buying?
\item \textbf{RQ2}: Can the specifically designed components of our model contribute to the performance improvement of group buying recommendation?
\item \textbf{RQ3}: How do the hyper-parameters of our model affect the performance of group buying recommendation?
\end{itemize}

\subsection{Experiment Data}
\subsubsection{Dataset of Group Buying}
By now, seldom group buying datasets from the real-world e-commerce platforms are available. The datasets used in the previous group recommendation models \cite{AttGroup,KgGroup,HierGroup} can not be used to evaluate our task of group buying, due to the essential difference between them that we have mentioned in Section \ref{sec:intro}. In our experiments, we used a public group buying dataset extracted from Beibei\footnote{https://github.com/Sweetnow/group-buying-recommendation} \cite{zhang2021group} to evaluate all compared models. Beibei is the largest e-commerce platform in China for maternal and infant products. This dataset contains the logs of sufficient group buying records, from which the initiator, product and participants of each deal group can be identified. Furthermore, two users (either an initiator or a participant) in a deal group of training set are recognized as a pair of social friends.

Although we only used this one dataset for our evaluations which only includes the products of maternal and infant, we believe that the evaluation results can justify our model's feasibility to other group buying scenarios, since the group buying on other e-commerce platforms has the same (or very similar) properties as that of Beibei.


\begin{table}[!htb]
\caption{Statistics of the preprocessed experiment dataset.}\label{tbl1}
\centering
\begin{tabular}{|c|l|}
\hline 

\hline
\textbf{Object} &  \textbf{Number}\\
\hline

\hline
user &  125,012\\
\hline item &  30,516\\
\hline deal group & 430,360 \\
\hline

\hline
\end{tabular}
\end{table}

\subsubsection{Sample Collection}
Before collecting the samples of our training and test set, we first filtered out the users who have less than five purchase records in the original Beibei dataset, which follows the popular manipulation in previous work \cite{BPR,FMF}. Then, we removed each group including the filtered users (no matter initiator or participant) and used the rest dataset in our experiment. The statistics of our preprocessed dataset is listed in Table \ref{tbl1}. 

The training and test samples of task A and task B were collected as follows. 
Each pair $(u, i)$ and triple $(u,i,p)$ observed from any dealt group in the processed dataset are used as one positive sample of task A and task B, respectively. Then, for any initiator $u$ of a group, we randomly selected a product (negative item) from the products that $u$ has not bought, to accompany $u$ as a negative sample of task A. For an observed deal group $<u, i, G>$ where $G$ is the participant set, we randomly selected a user (negative participant) from $\mathcal{U}\setminus G$ to accompany $(u, i)$ as a negative sample of task B. The ratio of positive samples to negative samples is 1:9. Accordingly, for each test instance in task A ($u$) and each test instance in task B ($(u, i)$), the sizes of their candidate lists are both 10. The model should compute the score for each item/participant in the candidate list, based on which each test instance's ranking list is determined. In addition, the ratio of training, validation and test set is 7:3:1. 


\subsection{Baselines}
We compared our model with the following recommendation models in our experiments to answer RQ1. In fact, all baselines can achieve the recommendation task of group buying through appropriate tailor although none of them were proposed specially for the two sub-tasks in this paper.



\textbf{DeepMF} ~\cite{xue2017deep}: It is a deep version of MF-based recommendation model, in which the rows and columns of user-item interaction matrix are input into multi-layer non-linear projection neural networks, to learn the latent representations of users and items.



\textbf{NGCF}~\cite{NGCF}: It is a deep CF model with GCNs, which explicitly models higher order connectivity between user-item interactions to improve embedding representations.

\textbf{DiffNet}~\cite{wu2019neural}: We believe that the social-based recommendation models enhancing user preference modeling with social links can also achieve the recommendation task of group buying, since we regard the relationships between initiators and participants as social links. Thus we also included social-based recommendation models into our comparisons. DiffNet is a representative social recommendation model with GCNs. It utilizes social relations and stimulates the social diffusion process on social networks for better representation learning of users and items. We consider this recommendation model in our comparisons due to that, the co-occurrence relationships between users observed from historical purchase records can be recognized as social links that are crucial for group buying recommendation.

\textbf{EATNN}~\cite{EATNN}: It is also a social-aware recommendation model that uses attention mechanisms to automatically assign personalized migration solutions to users by taking into account both their preferences on items and social relationships.

\textbf{GBGCN}~\cite{zhang2021group}: It is in fact one of group recommendation models. Although this model was proposed specifically for group buying and also distinguishes the roles of initiator and participant, it only achieves task A according to our task formalization. In this model, both user-item interaction graph and user-user social graph are constructed to learn the embeddings of users and items. An embedding propagation network is leveraged to extract user preferences in different roles.

\textbf{GBMF}~\cite{zhang2021group}: It was proposed as the variant of GBGCN, which directly uses dot-based similarity function to calculate scores of candidate items and candidate users as MF-based recommendation models. It also updates the embeddings of users and items for better performances.

We explain how to tailor these baselines to simultaneously achieve the two sub-tasks of group buying recommendation as our model. At first, task A can regarded as general item recommendation, all baselines can achieve it directly. Although all baselines were not designed specifically for task B, we can tailor them to achieve task B, only through altering their prediction layers and without modifying their major frameworks. In fact, the representations (embeddings) of $u, i, p$ are both learned in all of the baselines. According to their modeling principle, either embedding of $u, i, p$ has encoded the significant information from the rest two objects. Given that task B is similar to user recommendation \cite{jamali2010matrix,UserRec,kutty2014people}, i.e., for a given user ($u$) recommending another user ($p$) who has the same/similar preference as $u$ and would like to join $u$'s group, we can directly use the distance of $p$'s embedding and $u$'s embedding as $s(p|u,i)$. Specifically, we used inner product of two embeddings to measure their distance, since it is widely used in many recommendation models \cite{NCF,NGCF,lightGCN}.

In addition, we propose the following ablated variants of our MGBR to answer RQ2.

    \textbf{MGBR-M}: In this variant, the shared information of multi-task learning framework is removed from our model. In other words, expert network S and gate S are both removed, making our model degrade into a two-tower model over the three views. 
    
     \textbf{MGBR-R}: In this variant, we remove the auxiliary loss $\mathcal{L}'_A$ and $\mathcal{L}'_B$ in our model training.
     
    \textbf{MGBR-M-R}: In this variant, the shared information in the multi-task learning framework, as well as the two auxiliary losses are both removed.  
    
    \textbf{MGBR-G}: This variant replaces the adjusted gated units with generic gated units, indicating that $\mathbf{g}_{A2}^l$ in Eq. \ref{eq:gA} and $\mathbf{g}_{B2}^l$ in Eq. \ref{eq:gB} are both removed. In other words, $\alpha_A=\alpha_B=0$.
    
    \textbf{MGBR-D}: In this variant, the three divided views $\mathcal{G}_{UI},\mathcal{G}_{PI},\mathcal{G}_{UP}$ are replaced with an overall heterogeneous information network (HIN) including all nodes of $u, i, p$ and their relations. The embeddings of $u, i, p$ are learned by the GCN over this big HIN.

\begin{table}[!htb]
\centering
\caption{Some hyper-parameter settings in our experiments.}\label{tb:hyper}
{
\begin{tabular}{|c|c|l|}
\hline

\hline
\textbf{Para.}  & \textbf{Value} & \textbf{Comment}\\ 
 \hline
        
\hline
$d$ & 128 & embedding dimension   \\
\hline
$H$ &2 & the number of GCN layers\\
\hline
$K$ & 6 & the number of expert networks in each layer \\
\hline
$L$ & 2 & the layer number of expert networks and gates\\
\hline
$|\mathcal{T}|$ & 99 & negative sampling size in the auxiliary losses\\
\hline
$\alpha_A$ & 0.1 & control coefficient of Eq. \ref{eq:gA}\\
\hline
$\alpha_B$ & 0.1 & control coefficient of Eq. \ref{eq:gB}\\
\hline
$\beta$ & 1 & control coefficient of $\mathcal{L}_B$ in Eq. \ref{eq:L1}\\
\hline
$\beta_A$ & 0.3 & control coefficient of $\mathcal{L}'_A$ in Eq. \ref{eq:L1}\\
\hline
$\beta_B$ & 0.3 & control coefficient of $\mathcal{L}'_B$ in Eq. \ref{eq:L1}\\
\hline
$\rho$ & 0.0002 & learning rate\\
\hline
$|B|$ &64 & batch size\\
\hline

\hline
\end{tabular}
}
\end{table}

\begin{table*}[t]
    \label{tab:baselines}
    \centering
\caption{Overall performance comparisons between our MGBR and the baselines on task A and task B of group buying recommendation. It shows that MGBR's performance superiority over the baselines is more remarkable in task B.}\label{tb:overall}
\vspace{-0.2cm}
  {
    \begin{tabular}
     {|c|cc|cc|cc|cc|}
       \hline
       
       \hline
        \multirow{3}{*} {\diagbox{\textbf{Model}}{\textbf{Task}}} &\multicolumn{4}{c|}{\textbf{Task A}}&\multicolumn{4}{c|}{\textbf{Task B}} \\ 
        \cline{2-9}
       &\multicolumn{2}{c|}{\textbf{1:9}}&\multicolumn{2}{c|}{\textbf{1:99}}&\multicolumn{2}{c|}{\textbf{1:9}}&\multicolumn{2}{c|}{\textbf{1:99}}
        \\ 
         \cline{2-9}
          &MRR@10 &NDCG@10 &MRR@100 &NDCG@100 &MRR@10 &NDCG@10 & MRR@100 & NDCG@100 \\ 
                \hline
                
                \hline
       DeepMF     & 0.3763 & 0.5183 & 0.1672 & 0.3046 & 0.3070 & 0.4656 & 0.0654 & 0.2209\\
       \hline
        NGCF       & {0.5607} & {0.6617} & \underline{0.2841} & \underline{0.4150} & \underline{0.3778} & \underline{0.5211} & \underline{0.1254} & \underline{0.2748} \\
        \hline
        DiffNet      & 0.3780 & 0.5206 & 0.1290 & 0.2771 & 0.3314 & 0.4844 & 0.0976 & 0.2483\\
        \hline
        EATNN      & {\underline{0.5827}} & {\underline{0.6807}} & {0.2240} & {0.3736} & {0.3404} & {0.4929} & {0.0727} & {0.2310} \\
\hline
       GBGCN      & 0.5095 & 0.6231 & 0.2775 & 0.4006 & 0.3668 & 0.5127 & 0.1168 & 0.2665\\
       \hline
         GBMF      & 0.3718 & 0.5135 & 0.1433 & 0.2867 & 0.3254 & 0.4794 & 0.0884 & 0.2406\\
\hline
       \textbf{MGBR} & \textbf{0.6401} & \textbf{0.7292} & \textbf{0.2876} & \textbf{0.4501} & \textbf{0.6484} & \textbf{0.7327} & \textbf{0.2877} & \textbf{0.4471} \\
\hline
 Improvement   &{9.85\%}& {7.13\%}& 1.23\%& 8.46\%& 71.65\%& 40.61\%& 129.43\%& 62.70\%\\ 
         \hline
          
          \hline
    \end{tabular}
    }
    \vspace{-0.2cm}
\end{table*}

\subsection{Implementation Details} 
All of our experiments were conducted on the workstations of GeForce RTX 3090 with 24G memory and the environment of Ubuntu18.04.5 and torch1.8.0. The hyper-parameter settings of all baselines for the following displayed results were decided based on our tuning studies. For our MGBR, some hyper-parameters’ optimal values in the subsequent
experiments are listed in Table \ref{tb:hyper}, wherein the tuning study results of $\alpha_A, \alpha_B, \beta_A, \beta_B$ will be displayed afterwards.

\subsection{Evaluation Protocols}
In our experiments, We adopted \textbf{MRR@N} (mean reciprocal rank) and \textbf{NDCG@N} (normalized discounted cumulative gain) \cite{KTGAN,KAUR} to evaluate all models' recommendation performance, since they are very popular to evaluate ranking-based recommendation. 
\nop{
Specifically, MRR@N is the reciprocal rank of the only positive item/participant in the top-N ranking list averaged over all test instances. For a test instance, its NDCG@N is calculated as
\begin{equation}
\label{eq:nDCG}
NDCG@N=\frac{1}{Z}\sum_{i=1}^N\frac{2^{rel(i)}-1}{log_2(i+1)},
\end{equation}
where $rel(i)=1$ if the $i$-th item/participant in the ranking list is positive, otherwise 0, and $Z$ is the normalized factor.}
Particularly, NDCG is more sensitive than MRR to the positive item/participant's position in the ranking list. f
About the ratio of positive sample to negative sample in the test set, we also adopted 1:9 to compute MRR/NDCG@10, and further adopted 1:99 to compute MRR/NDCG@100 for all compared models.

\begin{table*}[t]
  \centering
 \caption{Performance comparisons between our model and the ablated variants.} \label{tb:ablation}
 \vspace{-0.2cm}
    \begin{tabular}{|c|cl|cl|cl|cl|}
    \hline

\hline
    \multirow{3}{*} {\diagbox{\textbf{Model}}{\textbf{Task}}}  & \multicolumn{8}{c|}{\textbf{Task A}} \\
     \cline{2-9}
    & \multicolumn{4}{c|}{\textbf{1:9}} & \multicolumn{4}{c|}{\textbf{1:99}} \\
      \cline{2-9}
   & \multicolumn{1}{l}{MRR@10} & \multicolumn{1}{l|}{R. Drop} & \multicolumn{1}{l}{NDCG@10} & \multicolumn{1}{l|}{R. Drop} & \multicolumn{1}{l}{MRR@100} & \multicolumn{1}{l|}{R. Drop} & \multicolumn{1}{l}{NDCG@100} & \multicolumn{1}{l|}{R. Drop} \\
     \hline
    MGBR-M-R & 0.2531 & -152.90\% & 0.4327 & -68.5\% & 0.0809 & -255.5\% & 0.2571 & -75.1\% \\
     \hline
    MGBR-M & 0.2607 & -145.53\% & 0.4401 & -65.7\% & 0.1217 & -136.3\% & 0.3095 & -45.4\% \\
     \hline
    MGBR-G & \underline{0.6126} & \underline{-4.49\%} & \underline{0.7041} & \underline{-3.56\%} & \underline{0.2732} & \underline{-5.27\%} & \underline{0.4322} & \underline{-4.14\%} \\
     \hline
    MGBR-R & 0.4228 & -51.40\% & 0.5663 & -28.77\% & 0.1221 & -135.54\% & 0.3136 & -43.53\% \\
     \hline
MGBR-D & 0.5189 & -23.36\% & 0.6390 & -14.12\% & 0.2091 & -37.54\% & 0.3793 &  -18.67\% \\
     \hline
    \textbf{MGBR} & \textbf{0.6401} & \textbf{--} & \textbf{0.7292} & \textbf{--} & \textbf{0.2876} & \textbf{--} & \textbf{0.4501} & \textbf{--} \\
     \hline

\hline
    & \multicolumn{8}{c|}{\textbf{Task B}} \\
     \hline
    MGBR-M-R & 0.2344 & -176.62\% & 0.4141 & -76.9\% & 0.1043 & -175.8\% & 0.2946 & -51.8\% \\
     \hline
    MGBR-M & 0.2471 & -162.40\% & 0.4272 & -71.5\% & 0.1147 & -150.8\% & 0.3051 & -46.5\% \\
     \hline
    MGBR-G & 0.4707 & -37.75\% & 0.6001 & -22.10\% & \underline{0.1797} & \underline{-60.10\%} & \underline{0.3448} &\underline{ -29.67\%} \\
     \hline
    MGBR-R & \underline{0.4769} & \underline{-35.96\%} & \underline{0.6074} & \underline{-20.63\%} & 0.1661 & -73.21\% & 0.3437 & -30.08\% \\
     \hline
MGBR-D & 0.4494 & -44.28\% & 0.5858 & -25.08\% & 0.1501 & -91.67\% & 0.3301 & -35.44\% \\
     \hline
    \textbf{MGBR} & \textbf{0.6484} & \textbf{--} & \textbf{0.7327} & \textbf{--} & \textbf{0.2877} & \textbf{--} & \textbf{0.4471} & \textbf{--} \\
        \hline

\hline
    \end{tabular}%
  \label{tab:addlabel}%
  \vspace{-0.2cm}
\end{table*}%

\subsection{Overall Performance Comparisons}
The performances scores of all compared models on the two sub-tasks are listed in Table \ref{tb:overall}, where the best scores are bold. As well, the relative performance improvements between our MGBR and the strongest baselines (highlighted with underline) are also displayed. All reported performance scores of each model are the average results of three runnings. 
Based on these results, we have the following observations and analyses.

 1. Obviously, our MGBR has the best performance in both of the two sub-tasks, which should be attributed to several factors. At first, thanks to the GCNs on the multiple views, MGBR can extract rich significant features from the correlations between initiators, items and participants. Second, the significant information is shared by the two sub-tasks in the multi-task learning module. Furthermore, the auxiliary losses help our model promote information interactions and reduce conflicts between the two sub-tasks. 
It is no surprise that our MGBR outperforms the baselines in task B more apparently than in task A, since none of the baselines are designed for task B. In addition, MGBR leverages the information of the given $u$ and $i$ more sufficiently to predict $p$ in task B.

 2. Our MGBR outperforms GBGCN and GBMF on the two sub-tasks, although these two baselines were also proposed for group buying. Compared with GBMF, GBGCN has better performance and further utilizes the features from the heterogeneous graph of initiators, items and participants, which also verifies the advantage of introducing graph structural feature and GCNs.

 3. DiffNet performs poorly, although it has excellent performance in social recommendation. The reason may be that the co-occurrence relationships of users in the same groups are not real social relationships, that only reflect the common preferences of users. Comparatively, EATNN performs better than DiffNet in Task A, but it is still inferior to our MGBR.

 4. The power of GCN is the major reason of NGCF's superior performance over other baselines. In addition, unlike DiffNet and GBGCN, it has no special design for capturing social context, avoiding the influence of fake social relationships in the dataset.

 5. All baselines fail to obtain satisfactory performance in task B, although all of them were trained on both task A and task B simultaneously. It implies that the training for task B is more difficult and worthwhile to be explored further. It is also necessary to design specific module for task B in general recommendation models.


\subsection{Ablation Study}
We also compared our MGBR with its ablated variants to answer RQ2. All compared variants' performance scores and their relative performance drops compared with MGBR are listed in Table \ref{tb:ablation}, based on which we have the following observations and analyses.

 1. MGBR's performance improvements over MGBR-M are more than that over MGBR-R and MGBR-G, showing that the shared expert networks and gates (S) are more significant than the auxiliary losses and adjusted gated units on improving our model's performance.

 2. MGBR's superiority over MGBR-R, and MGBR-M's superiority over MGBR-M-R both justify the effectiveness of refining representation learning with the auxiliary losses.



3. MGBR-G's performance drop relatively to MGBR on task B is more prominent than on task A. It is because task B is indeed more difficult, since it requires models to capture not only users' personalized preferences, but also the common preferences among different users. The adjusted gates in MGBR was proposed to better achieve to the multi-task learning including the auxiliary learning objectives, which leverages prior experiences ($\mathbf{e}_u||\mathbf{e}_i, \mathbf{e}_i||\mathbf{e}_p, \mathbf{e}_u||\mathbf{e}_p$) to assign appropriate weights to different experts. Compared with task B, task A is less sensitive to the choice of experts since it is more simple.
 
 4. MGBR-D is also inferior to MGBR, showing that including all user and item nodes, along with their relations into an HIN can not obtain the optimal node embeddings through the GCN over the HIN. In fact, the divided views in MGBR can effectively alleviate the negative interactive effects between different relations.

\begin{table}[t]
  \centering
  \caption{Comparisons of model scale and time consumption (minutes).}\label{tb:time}
   \vspace{-0.2cm}
    \begin{tabular}{|c|r|c|}
    \hline

\hline
    \multicolumn{1}{|c|}{\textbf{Model}} & \multicolumn{1}{c|}{\textbf{Para. number}} & \multicolumn{1}{c|}{\textbf{Min./epoch}} \\
    \hline

\hline
  DeepMF & 155,500 & 0.34 \\
    \hline
  NGCF  & 9,962,176 & 3.17 \\
    \hline
    DiffNet & 15,556,217 & 1.67 \\
    \hline
    EATNN & 33,966,534 & 1.23 \\
    \hline
    GBGCN & 15,555,273 & 1.79 \\
    \hline
     GBMF  & 1,555,280 & 1.03 \\
    \hline
    MGBR  & 31,341,038 & 8.35 \\
    \hline

\hline
    \end{tabular}%
  \label{tab:addlabel}%
    \vspace{-0.2cm}
\end{table}%

\vspace{-0.4cm}
\subsection{Model Scale and Efficiency Comparisons}
Table \ref{tb:time} lists the scale of model parameters and the time consumption (minutes) of one training epoch for all compared models. As we mentioned before, our MGBR is the most time-consuming as it has more complex architecture (more parameters) than the baselines. In EATNN, each user is represented by three kinds of embeddings, so it even has more parameters than our MGBR due to its great number of users. However, it only adopts attentions and MLP operations which are relatively simple compared with MGBR's architecture, thus consumes less time.

\subsection{Influence of Important Hyper-parameters}
To answer RQ3, we also investigated the influence of some important hyper-parameters of our model on recommendation performance.
\subsubsection{Impact of Auxiliary Loss Weight}
\begin{figure}[t]
  \centering
  \includegraphics[width=0.98\columnwidth]{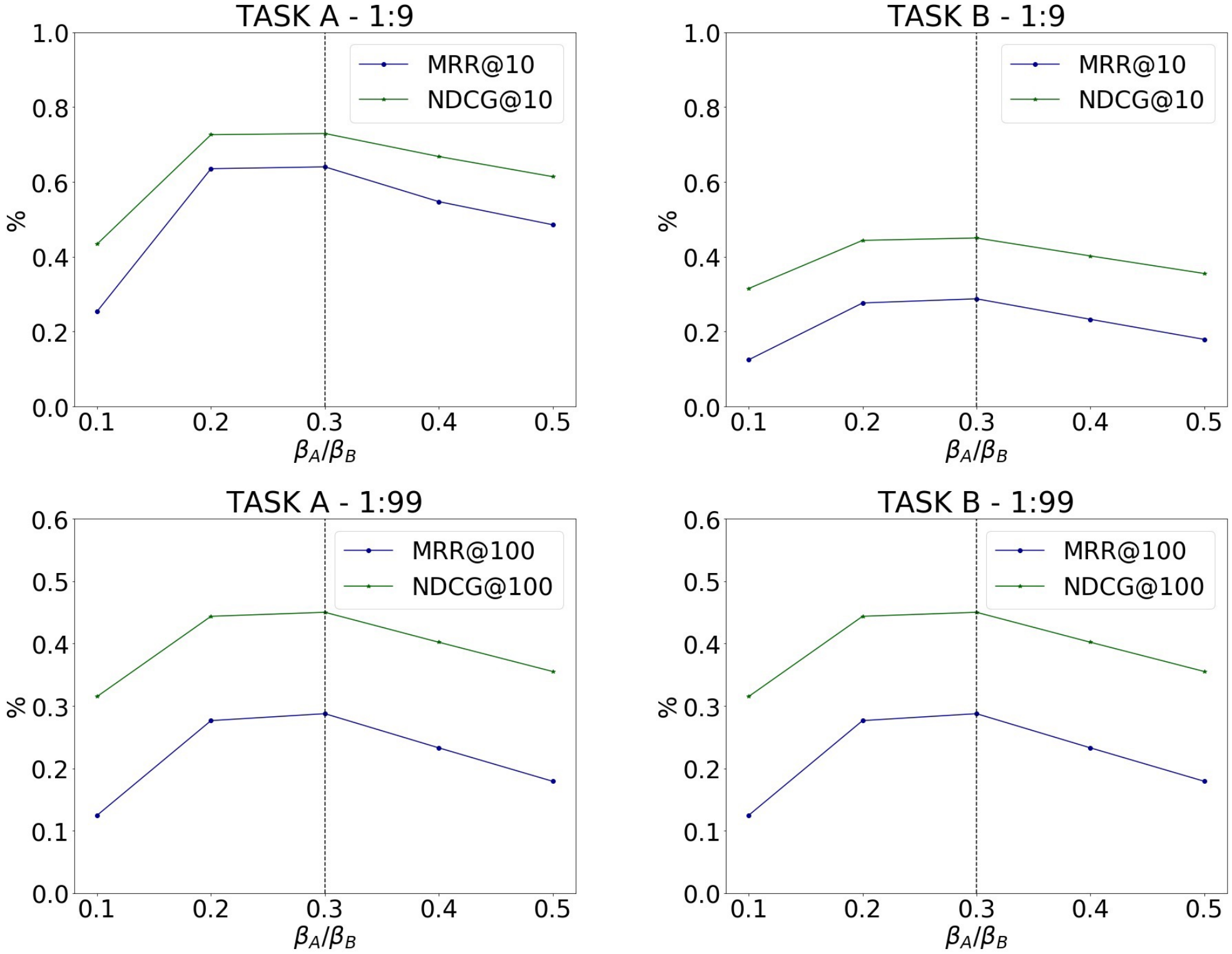}
  \vspace{-0.2cm}
  \caption{MGBR's performance with different auxiliary loss weights.}\label{fig:beta}
  5\vspace{-0.1cm}
\end{figure}

The ablation studies have verified that the two auxiliary losses, i.e., $\mathcal{L}'_A$ and $\mathcal{L}'_B$ are helpful for our model to obtain improved performance. To analyze the influence of the weight of $\mathcal{L}'_A$ and $\mathcal{L}'_B$ we set $\beta_A=\beta_B$ in Eq. \ref{eq:L1} and set their values in the range of \{0.1, 0.2, 0.3, 0.4, 0.5\}, and recorded the corresponding recommendation performance comparison on the two sub-tasks as shown in Fig. \ref{fig:beta}. According to the tuning results, MGBR performs the best when $\beta_A=\beta_B=0.3$. It shows that either lower or higher value of $\beta_A$ and $\beta_B$ would hurt model performance. 
Too small $\beta_A$ and $\beta_B$ make it difficult to constrain our model with the auxiliary losses and reduce the model's generalization. While too large $\beta_A$ and $\beta_B$ would cause our model to overlook fitting with the observed group buying records.

\subsubsection{Impact of Control Coefficient in Adjusted Gate}
\begin{figure}[t]
  \centering
  \includegraphics[width=0.98\columnwidth]{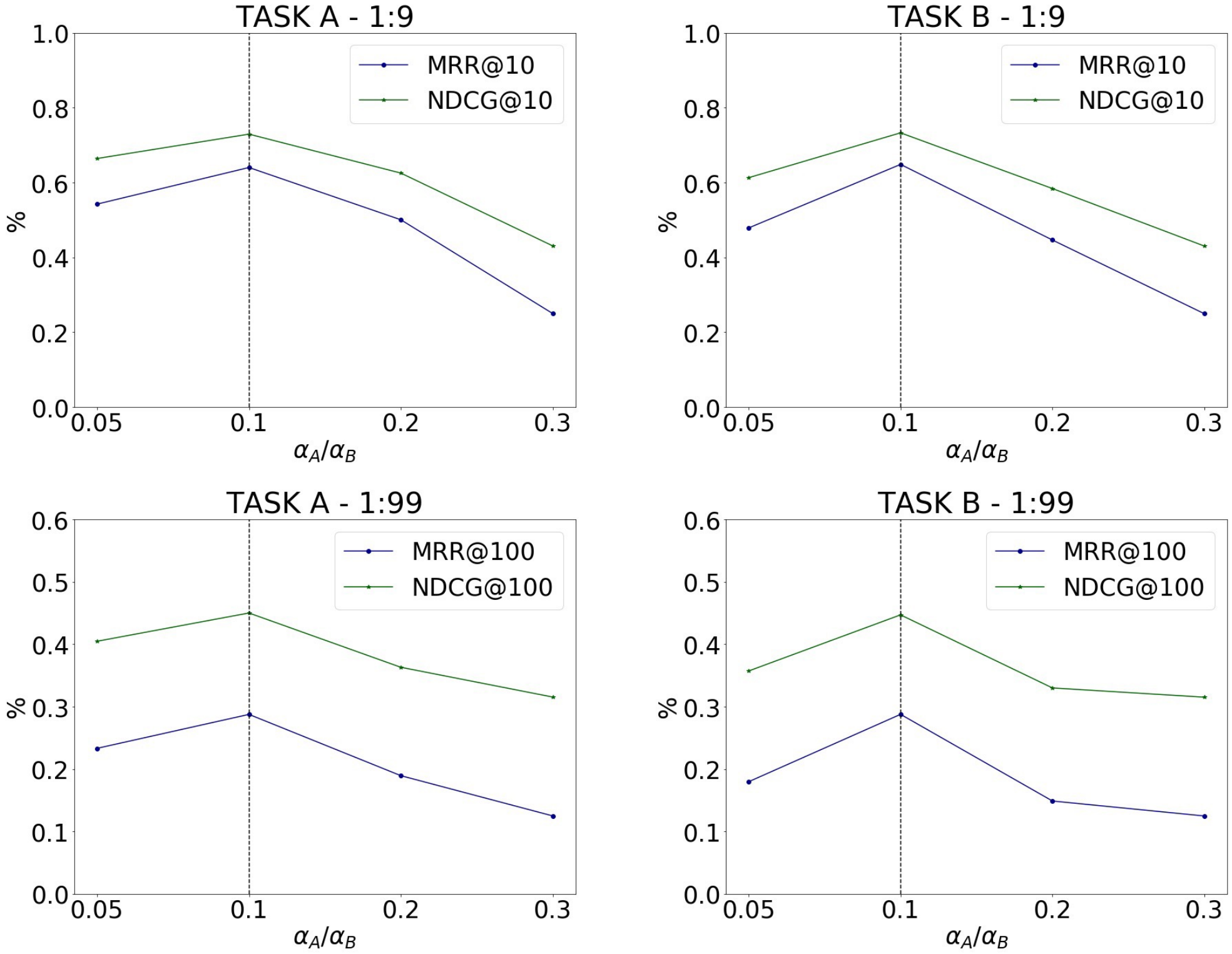}
  \vspace{-0.2cm}
  \caption{MGBR's performance with different control coefficients of adjusted gates.}\label{fig:alpha}
  \vspace{-0.2cm}
\end{figure}

To analyze the influence of the control coefficient in the adjusted gates, we also set $\alpha_A$ in Eq. \ref{eq:gA} the same as $\alpha_B$ in Eq. \ref{eq:gB}, and varied their values in the range of \{0.05, 0.1, 0.2, 0.3\}. Our model's performance corresponding to these values are displayed in Fig. \ref{fig:alpha}, showing that $\alpha_A=\alpha_B=0.1$ is the optimal setting for our model to achieve the best performance. The large $\alpha_A$ and $\alpha_B$ indicate that the gate information is extracted in terms of the information of $(u,i,p)$ more than the expert networks, resulting in the insufficient utilization of expert network's information. The small $\alpha_A$ and $\alpha_B$ result in the insufficient utilization of $(u, i, p)$'s information, and also hurt the model's performance.

\subsection{Case Study of Representation Learning}
As other recommendation models, the representation learning of initiators, items and participants is the key of our model to achieve the task of group buying recommendation. Optimal object embeddings in the model evidently result in better recommendation performance. In this subsection, we visualize the effectiveness of the shared information (expert network and gate S) in our multi-task learning module and the auxiliary losses on refining the object embeddings through some cases.
\begin{figure}[t]
 \centering
  \includegraphics[width=0.97\linewidth]{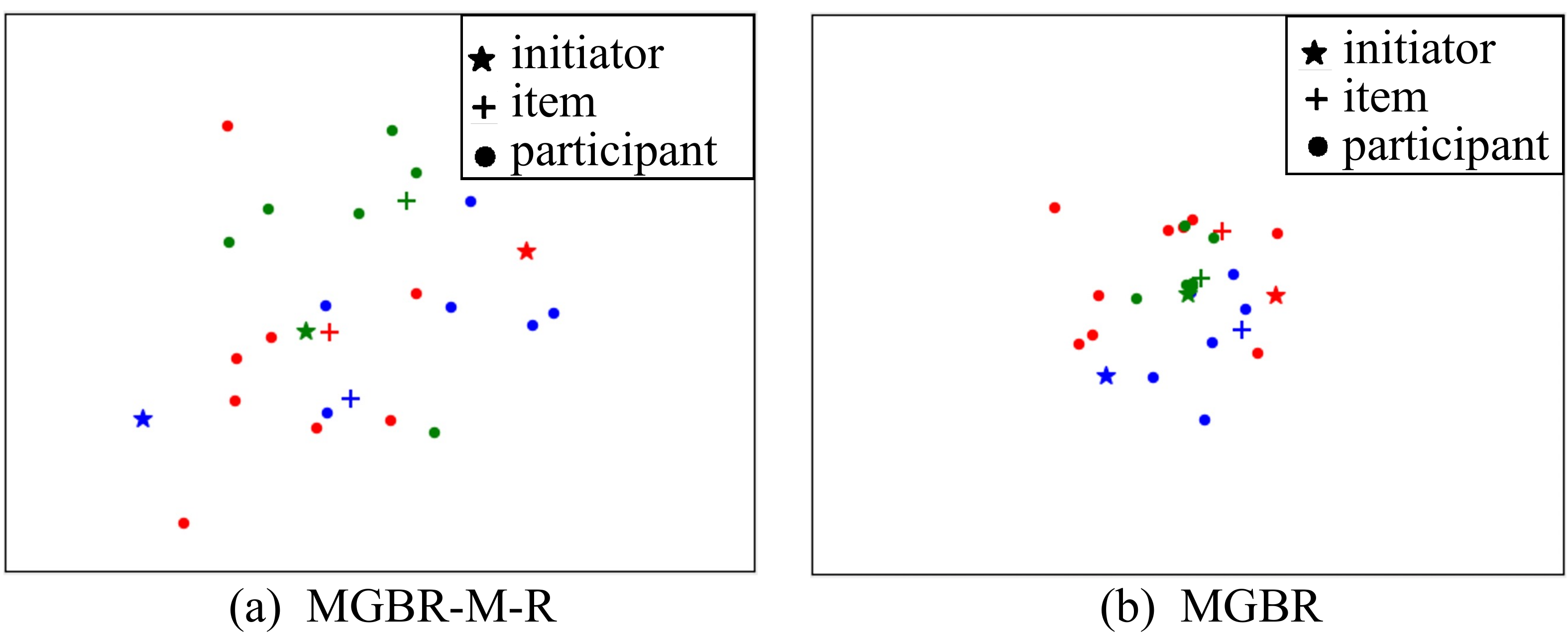}
  \vspace{-0.2cm}
   \caption{The object embedding comparisons between MGBR and MGBR-M-R for some cases (better viewed in color). It shows that the shared information in the multi-task learning module and the two auxiliary losses help our model learn better embeddings of initiators, items, participants.}\label{fig:emb}
\vspace{-0.2cm} 
\end{figure}

To this end, we applied PCA decomposition to project all object embeddings into a 2-D space. Fig. \ref{fig:emb} displays the embeddings of some groups in MGBR-M-R (a) and MGBR (b), where the colors indicate different groups. The stars, plusses and dots represent the embeddings of initiators, items and participants, respectively. From the figure, we observe that the users and items of the same group (the same color) in MGBR are more concentrated than that in MGBR-M-R, and the embeddings of different groups are distinguishable. It shows that, with the shared information in the multi-task learning module and auxiliary losses, our model is capable of leveraging the information interactions within one group to learn better object embeddings, and thus improves recommendation performance. 


\section{Related Work}\label{sec:related}

\subsection{Recommender Systems}
\subsubsection{Item Recommendation}
Item recommendation~\cite{sarwar2001item} aims to predict the probability of a user interacting with a candidate item, and then recommend the item with the highest probability from the candidate set to the user. In recent years, many deep recommendation models \cite{NCF,xue2017deep,KTGAN,KAUR} have been proposed. Besides leveraging user-item interactions to infer the relationships between user and items, some literatures have introduced social context into item recommendation, based on the assumption that a user has the same or similar preference as his/her friends. Jamali et al.~\cite{jamali2010matrix} introduced social relations in MF~\cite{koren2009matrix} to make the representations of social friends more close, to enhance recommendation performance. \cite{davoudi2017effects} recommends items for users through combining a user's preference and his friends' preferences. In addition, \cite{wu2019neural} uses the propagation of user embedding on social networks to capture social influences, and \cite{EATNN} adopts attention mechanisms to automatically assign personalized migration solutions to users by considering both their preferences on items and social relationships.


\subsubsection{User Recommendation}
Besides item recommendation, user recommendation~\cite{kutty2014people,UserRec} is another popular scheme in recommender systems, which generally predicts the probability of the interaction between two users. 
Traditional user recommendation task is just a binary classification to judge whether a candidate user deserves to be recommended to the target user. The second sub-task in our group buying recommendation is in fact a user recommendation task, but the given object is an item and an initiator instead of a sole user. 
The classical recommendation methods such as MF can also be used to achieve user recommendation. For example, SocialMF~\cite{jamali2010matrix} adds trust propagation to MF, so as to improve recommendation performance. SocialReg~\cite{ma2011recommender} uses two kinds of social information to design social regularization terms to constraint matrix decomposition objective function. Besides the social links between users \cite{SocialRec}, other features about user preferences can also be leveraged in promote recommendation performance, including user tags \cite{SocialTag}, activities \cite{UserRec}, etc.

\subsubsection{Group Recommendation}
Although there is essential difference between group recommendation \cite{Group1,Group} and group buying recommendation, group recommendation is still the recommendation category most related to group buying recommendation. 
The key problems to achieve group recommendation include how to fuse the preference of each user in a group into the group's preference \cite{Group2}, and how to handle the disagreements among group members \cite{Group1}. The classical approaches include~\cite{baltrunas2010group,yuan2014generative,quintarelli2016recommending}. Cao et al.~\cite{AttGroup} used attention mechanism to learn precious user and group interest representations. Since the social relationships among users indicate their preference similarities, many group recommendation models leverage social information to learn the group representations \cite{SocialGroup,yin2019social,cao2019social,SocialGR}. Specifically, GBGCN~\cite{zhang2021group} was proposed to capture structural information among group buying records, and model user dual role, social influence and complex group buying implicit feedback. Although it also focuses on group buying, it still aims to solve the first sub-task, i.e., recommending items to a given group as other group recommendation models. For occasional group recommendation, GroupSA \cite{GroupSA} models the group decision making process as multiple voting processes, and uses a stacked social self-attention network to simulate how a group consensus is reached. In addition, the knowledge graph (KG) is also helpful to learning group representations \cite{KgGroup}. Besides online shopping, group recommendation is also applied in many other areas, such as POI recommendation~\cite{zhu2018context} and restaurant recommendation~\cite{zhu2020context}.  

\subsubsection{Graph Neural Networks for Recommendation}
Graph neural networks (GNNs) can capture structural features of graphs via message passing between the nodes, which have been widely used in recommender systems. For example, GNN is used in the matrix completion task for recommendation~\cite{berg2017graph}. NGCF~\cite{NGCF} uses embedding propagation on the bipartite graph to model high-order collaborative signals. In addition, GNNs have been successfully employ in session-based recommendation. Wu et al.~\cite{wu2019session} used GNNs to capture item relation between different sessions and in knowledge-based recommendation. KGAT~\cite{wang2019kgat} harnesses graph attention network (GAT) \cite{GAT} to obtain more accurate and sufficient feature representations. Meng et al. \cite{MKM-SR} leveraged GGNN \cite{GGNN} to capture the complex patterns among the items in a session.

\subsection{Multi-task Learning}
Multi-task learning (MTL) helps recommender systems achieve better recommendation through leveraging the information from different aspects, which has been widely incorporated into many recommender systems \cite{MKM-SR}. A classical MTL approach is hard-parameter sharing, forcing different tasks to share the same parameters. For example, ~\cite{caruana1997multitask} uses the shared-bottom structure to make all tasks share the bottom layer parameters. In industry, ESSM \cite{ma2018entire} achieves CTR and CVR tasks simultaneously, and generates outstanding results on large-scale data. Similarly, ~\cite{wang2022escm} adds a counterfactual learning task to ESSM to alleviate sample selection bias (SSB), and thus achieves better performance on industrial data.

Comparatively, soft-parameter sharing only requires to share a part of the parameters rather. Specifically, it allows each task to select a sub-network, reducing the impact of correlation between tasks. \cite{wang2022escm, shazeer2017outrageously} take mixture-of-expert as an important component of MTL, employing different experts to learn different aspects of the features. 
\cite{ma2018modeling} proposes the idea of multi-gate, that makes it possible for expert networks to pass different information to different tasks. 
Ma et al.~\cite{ma2019snr} proposed a more flexible soft-parameter sharing approach to add parameter sharing process inside the shared layers, and make the information propagation between expert networks trainable. \cite{tang2020progressive} explicitly divides the expert networks into task-share and task-specific parts, and replaces the fully connected propagation between expert layers with the propagation between the two parts. Accordingly, the MTL of our MGBR also belongs to soft-parameter sharing.

More recently, more flexible and robust MTL approaches have been proposed. For example, AITM~\cite{xi2021modeling} takes the sequential relationship into account and uses an information migration module to model task dependencies. In this way, information migration occurs not only at the bottom layers, but also at the layers close to the output. \cite{zou2022multi} introduces contrastive learning \cite{CL} to alleviate the parameter conflict problem in MTL. Sun et al.~\cite{sun2020learning} proposed the sparse-sharing approaches, in which the parameters are partially shared across tasks, making it more flexible to handle loosely related tasks.

\section{Conclusion}\label{sec:conclude}
In this paper, we pioneer the task formalization of group buying recommendation for the online shopping platforms. Aiming to achieve the two sub-tasks of group buying recommendation simultaneously, we propose a novel recommendation model MGBR. Based on our insights into the correlations and information interaction between the two sub-tasks, we design collaborative expert networks and adjusted gated units in MGBR's multi-task learning module, to better learn the object embeddings in buying groups. Furthermore, we propose two auxiliary losses corresponding to the two sub-tasks for model training, to refine representation learning. Our extensive experiments evidently justify MGBR's superior performance of group buying recommendation over the previous recommendation models.
\bibliographystyle{IEEEtran}
\bibliography{refer}	

\end{document}